\documentclass[12pt]{article}
\usepackage{amssymb,amsmath,epsfig}
\begin{document}
\title{\bf Anisotropic Dark Energy Bianchi Type $III$ Cosmological Models in Brans Dicke Theory of Gravity}

\author{M. Farasat Shamir \thanks{farasat.shamir@nu.edu.pk} and
Akhlaq Ahmad Bhatti \thanks{akhlaq.ahmad@nu.edu.pk}\\\\
Department of Sciences and Humanities,\\
National University of Computer and \\Emerging Sciences,
Lahore-54770, Pakistan.}

\date{}

\maketitle
\begin{abstract}
The main purpose of this paper is to explore the solutions of
Bianchi type $III$ cosmological model in Brans Dicke theory of
gravity in the background of anisotropic dark energy. We use the
assumption of constant deceleration parameter and power law
relation between scalar field $\phi$ and scale factor $a$ to find
the solutions. The physical behavior of the solutions has been
discussed using some physical quantities.
\end{abstract}


\section{Introduction}

Recent experimental data \cite{1} about late time accelerated
expansion of the universe has attracted much attention in the
recent years. Cosmic acceleration can be well explained from high
red-shift supernova experiments . The recent results from cosmic
microwave background fluctuations \cite{2} and large scale
structure \cite{3} suggest the expansion of universe. Dark energy
seems to be best candidate to explain cosmic acceleration. It is
now believed that $96$ percent energy of the universe consist of
dark energy and dark matter ($76$ percent dark energy and $20$
percent dark matter) \cite{2, 04}. Dark energy is the most popular
way to explain recent observations that the universe is expanding
at an accelerating rate. The exact nature of the dark energy is a
matter of speculation. It is known to be very homogeneous, not
very dense and is not known to interact through any of the
fundamental forces other than gravity. Since it is not very dense,
roughly $10-29$ grams per cubic centimeter, so it is a difficult
task to detect it in the laboratory. It is thought that dark
energy have a strong negative pressure in order to explain the
observed acceleration in the expansion rate of the universe.

Dark energy models have significant importance now as far as
theoretical study of the universe is concerned. It would be more
interesting to study the variable equation of state (EoS), i.e.
$P=\rho\omega(t)$, where $P$ is the pressure and $\rho$ is the
energy density of universe. Usually EoS parameter is assumed to be
a constant with the values $-1,~0,~-\frac{1}{3}$ and $+1$ for
vacuum, dust, radiation and stiff matter dominated universe,
respectively. However, it is a function of time or redshift
\cite{004} in general. Latest observations \cite{4} from SNe Ia
data indicate that $\omega$ is not constant. In recent years, many
authors \cite{5}-\cite{5005} have shown keen interests in studying
the universe with variable EoS. Sharif and Zubair \cite{5}
discussed the dynamics of Bianchi type $VI_0$ universe with
anisotropic dark energy in the presence of electromagnetic field.
The same authors \cite{5001} explored Bianchi type $I$ universe in
the presence of magnetized anisotropic dark energy with variable
EoS parameter. Akarsu and Kilinc \cite{5003} investigated the
general form of the anisotropy parameter of the expansion for
Bianchi type $III$ model.

The isotropic models are considered to be most suitable to study
large scale structure of the universe. However, it is believed
that the early universe may not have been exactly uniform. This
prediction motivates us to describe the early stages of the
universe with the models having anisotropic background. Thus, it
would be worthwhile to explore anisotropic dark energy models in
the context of modified theories of gravity. Among the various
modifications of general relativity (GR), the Brans-Dicke (BD)
theory of gravity \cite{6} is a well known example of a scalar
tensor theory in which the gravitational interaction involves a
scalar field and the metric tensor. One extra parameter $\varpi$
is used in this theory which satisfies the equation given by
\begin{equation*}
\Box \phi=\frac{8\pi T}{3+2\varpi},
\end{equation*}
where $\phi$ is known as BD scalar field while $T$ is the trace of
the matter energy-momentum tensor. It is mentioned here that the
general relativity is recovered in the limiting case
$\varpi\rightarrow \infty$. Thus we can compare our results with
experimental tests for significantly large value of $\varpi$.

Bianchi type models are among the simplest models with anisotropic
background. Many authors \cite{614}-\cite{619} explored Bianchi
type spacetimes in different contexts. Moussiaux et al. \cite{620}
investigated the exact solution for vacuum Bianchi type-$III$
model with a cosmological constant. Xing-Xiang \cite{621}
discussed Bianchi type $III$ string cosmology with bulk viscosity.
He assumed that the expansion scalar is proportional to the shear
scalar to find the solutions. Wang \cite{6022} investigated string
cosmological models with bulk viscosity in Kantowski-Sachs
spacetime. Upadhaya \cite{622} explored some magnetized Bianchi
type-$III$ massive string cosmological models in GR. Hellaby
\cite{6023} gave an overview of some recent developments in
inhomogeneous models and it was concluded that the universe is
inhomogeneous on many scales. Sharif and Shamir \cite{14,15} have
studied the solutions of Bianchi types $I$ and $V$ spacetimes in
the framework of $f(R)$ gravity. Recently, we \cite{16,17}
explored the exact vacuum solutions of Bianchi type $I$, $III$ and
Kantowski- Sachs spacetimes in the metric version of f(R) gravity.

The study of Bianchi type models in the context of BD theory has
attracted many authors in the recent years \cite{7}. A detailed
discussion of BD cosmology is given by Singh and Rai \cite{8}.
Lorenz-Petzold \cite{9} studied exact Bianchi type-$III$ solutions
in the presence of electromagnetic field. Kumar and Singh
\cite{10} investigated perfect fluid solutions using Bianchi type
$I$ spacetime in scalar-tensor theory. Adhav et al. \cite{11}
obtained an exact solution of the vacuum Brans-Dicke field
equations for the metric tensor of a spatially homogeneous and
anisotropic model. Pradhan and Amirhashchi \cite{12} investigated
anisotropic dark energy Bianchi type $III$ model with variable EoS
parameter in GR. Adhav et al. \cite{13} explored Bianchi type
$III$ cosmological model with negative constant deceleration
parameter in Brans Dicke theory of gravity in the presence of
perfect fluid.

In this paper, we focuss our attention to explore the solutions of
anisotropic dark energy Bianchi type $III$ cosmological model in
the context of BD theory of gravity. We find the solutions using
the assumption of constant deceleration parameter and power law
relation between $\phi$ and $a$. The paper is organized as
follows: A brief introduction of the field equations in BD theory
of gravity is given in section \textbf{2}. In section \textbf{3},
the solutions of the field equations for Bianchi types $III$
spacetime are found. Some physical and kinematical parameters are
also evaluated for the solutions. Singularity analysis is given in
section \textbf{4}. A brief summary is given in the last section.

\section{Some Basics of Brans Dicke theory of Gravity}

The line element for the spatially homogeneous and anisotropic
Bianchi type $III$ spacetime is given by
\begin{equation}\label{1}
ds^{2}=dt^2-A^2(t)dx^2-e^{-2mx}B^2(t)dy^2-C^2(t)dz^2,
\end{equation}
where $A,~B$ and $C$ are cosmic scale factors and $m$ is a
positive constant. The energy momentum tensor for anisotropic dark
energy is given by
\begin{equation}\label{2}
{T^i}_{j}=diag[\rho,-p_x,-p_y,-p_z]=diag[1,-\omega_x,-\omega_y,-\omega_z]\rho,
\end{equation}
where $\rho$ is the energy density of the fluid while $p_x,~p_y$
and $p_z$ are the pressures on the $x$, $y$ and $z$ axes
respectively. Here $\omega$ is EoS parameter of the fluid with no
deviation and $\omega_x$, $\omega_y$ and $\omega_z$ are the EoS
parameters in the directions of x, y and z axes respectively. The
energy momentum tensor can be parameterized as
\begin{equation}\label{3}
{T^i}_{j}=diag[1,-\omega,-(\omega+\gamma),-(\omega+\delta)]\rho.
\end{equation}
For the sake of simplicity, we choose $\omega_x=\omega$ and the
skewness parameters $\gamma$ and $\delta$ are the deviations from
$\omega$ on $y$ and $z$ axes respectively.

The Brans-Dicke field equations are
\begin{equation}\label{4}
R_{ij}-\frac{1}{2}Rg_{ij}-\frac{\varpi}{\phi^2}(\phi_{,i}\phi_{,j}
-\frac{1}{2}g_{ij}\phi_{,k}\phi^{,k})-\frac{1}{\phi}(\phi_{;ij}-g_{ij}\Box
\phi)=\frac{8\pi T_{ij}}{\phi},
\end{equation}
and
\begin{equation}\label{5}
\Box \phi=\phi^{,k}_{;k}=\frac{8\pi T}{3+2\varpi},
\end{equation}
where $\varpi$ is a dimensionless coupling constant. For Bianchi
type $III$ spacetime, the field equations take the form
\begin{eqnarray}\label{6}
\frac{\dot{A}\dot{B}}{AB}+\frac{\dot{B}\dot{C}}{BC}+
\frac{\dot{C}\dot{A}}{CA}-\frac{m^2}{A^2}-
\frac{\varpi}{2}(\frac{\dot{\phi}}{\phi})^2+
\frac{\dot{\phi}}{\phi}(\frac{\dot{A}}{A}+\frac{\dot{B}}{B}+
\frac{\dot{C}}{C})=\frac{8\pi\rho}{\phi},\\
\label{7} \frac{\ddot{B}}{B}+\frac{\ddot{C}}{C}+
\frac{\dot{B}\dot{C}}{BC}+\frac{\ddot{\phi}}{\phi}+
\frac{\varpi}{2}(\frac{\dot{\phi}}{\phi})^2+
\frac{\dot{\phi}}{\phi}(\frac{\dot{B}}{B}+
\frac{\dot{C}}{C})=-\frac{8\pi\omega\rho}{\phi},\\
\label{8} \frac{\ddot{C}}{C}+\frac{\ddot{A}}{A}+
\frac{\dot{C}\dot{A}}{CA}+\frac{\ddot{\phi}}{\phi}+
\frac{\varpi}{2}(\frac{\dot{\phi}}{\phi})^2+
\frac{\dot{\phi}}{\phi}(\frac{\dot{C}}{C}+
\frac{\dot{A}}{A})=-\frac{8\pi(\omega+\gamma)\rho}{\phi},\\
\label{9}\frac{\ddot{A}}{A}+\frac{\ddot{B}}{B}- \frac{m^2}{A^2}+
\frac{\dot{A}\dot{B}}{AB}+\frac{\ddot{\phi}}{\phi}+
\frac{\varpi}{2}(\frac{\dot{\phi}}{\phi})^2+
\frac{\dot{\phi}}{\phi}(\frac{\dot{A}}{A}+
\frac{\dot{B}}{B})=-\frac{8\pi(\omega+\delta)\rho}{\phi}.
\end{eqnarray}
Also, the $01$-component can be written in the following form
\begin{equation}\label{10}
\frac{\dot{A}}{A}- \frac{\dot{B}}{B}=0.
\end{equation}
Integrating this equation, we obtain
\begin{equation}\label{12}
B=c_1A,
\end{equation}
where $c_1$ is an integration constant. Without loss of any
generality, we take $c_1=1$. Using Eq.(\ref{5}), we get
\begin{equation}\label{11}
\ddot{\phi}+\dot{\phi}(\frac{\dot{A}}{A}+\frac{\dot{B}}{B}+
\frac{\dot{C}}{C})=\frac{8\pi(1-3\omega-\delta-\gamma)\rho}{\phi(3+2\varpi)}.
\end{equation}

Now we define some physical parameters before solving the field
equations.

The average scale factor $a$ and the volume scale factor $V$ are
defined as
\begin{eqnarray}\label{60-8}
a=\sqrt[3]{A^2C},\quad V=a^3=A^2C.
\end{eqnarray}
The generalized mean Hubble parameter $H$ is given in the form
\begin{equation}\label{60-008}
H=\frac{1}{3}(H_1+H_2+H_3),
\end{equation}
where $H_1=\frac{\dot{A}}{A}=H_2,~H_3=\frac{\dot{C}}{C}$ are the
directional Hubble parameters in the directions of $x,~y$ and $z$
axis respectively. Using Eqs.(\ref{60-8}) and (\ref{60-008}), we
obtain
\begin{equation}\label{60-0008}
H=\frac{1}{3}\frac{\dot{V}}{V}=\frac{1}{3}(H_1+H_2+H_3)=\frac{\dot{a}}{a}.
\end{equation}
The expansion scalar $\theta$ and shear scalar $\sigma$ are
defined as follows
\begin{eqnarray}\label{60-09}
\theta&=&u^\mu_{;\mu}=2\frac{\dot{A}}{A}+\frac{\dot{C}}{C},\\
\label{60-00009}
\sigma^2&=&\frac{1}{2}\sigma_{\mu\nu}\sigma^{\mu\nu}
=\frac{1}{3}[\frac{\dot{A}}{A}-\frac{\dot{C}}{C}]^2,
\end{eqnarray}
where
\begin{equation}\label{60-009}
\sigma_{\mu\nu}=\frac{1}{2}(u_{\mu;\alpha}h^\alpha_\nu+u_{\nu;\alpha}h^\alpha_\mu)
-\frac{1}{3}\theta h_{\mu\nu},
\end{equation}
$h_{\mu\nu}=g_{\mu\nu}-u_{\mu}u_{\nu}$ is the projection tensor
while $u_\mu=\sqrt{g_{00}}(1,0,0,0)$ is the four-velocity in
co-moving coordinates. The mean anisotropy parameter $A_m$ is
defined as
\begin{equation}\label{60-1008}
A_m=\frac{1}{3}\sum(\frac{\triangle H_i}{H})^2,
\end{equation}
where $\triangle H_i=H_i-H$, $(i=1,2,3)$.

\section{Solution of the Field Equations}
Subtracting Eq.(\ref{7}) and Eq.(\ref{8}), we get
\begin{equation}\label{13}
\frac{\ddot{B}}{B}-\frac{\ddot{A}}{A}+
\frac{\dot{C}}{C}(\frac{\dot{B}}{B}-\frac{\dot{A}}{A})+
\frac{\dot{\phi}}{\phi}(\frac{\dot{B}}{B}-\frac{\dot{A}}{A})=\frac{8\pi\gamma\rho}{\phi}.
\end{equation}
Using Eq.(\ref{12}), this equation gives $\gamma=0$. Thus
Eqs.(\ref{6}-\ref{9}) and Eq.(\ref{11}) reduce to
\begin{eqnarray}\label{14}
(\frac{\dot{A}}{A})^2+ \frac{2\dot{C}\dot{A}}{CA}-\frac{m^2}{A^2}-
\frac{\varpi}{2}(\frac{\dot{\phi}}{\phi})^2+
\frac{\dot{\phi}}{\phi}(2\frac{\dot{A}}{A}+
\frac{\dot{C}}{C})=\frac{8\pi\rho}{\phi},\\
\label{15} \frac{\ddot{C}}{C}+\frac{\ddot{A}}{A}+
\frac{\dot{C}\dot{A}}{CA}+\frac{\ddot{\phi}}{\phi}+
\frac{\varpi}{2}(\frac{\dot{\phi}}{\phi})^2+
\frac{\dot{\phi}}{\phi}(\frac{\dot{C}}{C}+
\frac{\dot{A}}{A})=-\frac{8\pi\omega\rho}{\phi},\\
\label{16}\frac{2\ddot{A}}{A}+(\frac{\dot{A}}{A})^2-
\frac{m^2}{A^2}+\frac{\ddot{\phi}}{\phi}+
\frac{\varpi}{2}(\frac{\dot{\phi}}{\phi})^2+
\frac{2\dot{\phi}\dot{A}}{\phi A}=-\frac{8\pi(\omega+\delta)\rho}{\phi},\\
\label{17}\ddot{\phi}+\dot{\phi}(\frac{2\dot{A}}{A}+
\frac{\dot{C}}{C})=\frac{8\pi(1-3\omega-\delta)\rho}{\phi(3+2\varpi)}.
\end{eqnarray}
Integration after subtracting Eq.(\ref{15}) and Eq.(\ref{16})
yields
\begin{equation}\label{18}
\frac{\dot{A}}{A}-\frac{\dot{C}}{C}=\frac{1}{A^2C\phi}
\int(\frac{m^2}{A^2}-\frac{8\pi\delta\rho}{\phi})\phi A^2C ~dt+
\frac{\lambda}{A^2C\phi},
\end{equation}
where $\lambda$ is an integration constant. The integral term in
this equation vanishes for
\begin{equation}\label{19}
\delta=\frac{m^2\phi}{8\pi\rho A^2}.
\end{equation}
Using Eq.(\ref{19}) in Eq.(\ref{18}), it follows that
\begin{equation}\label{20}
\frac{A}{C}=c_2e^{\lambda\int\frac{dt}{a^3\phi}},
\end{equation}
where $a^3=A^2C$ and $c_2$ is an integration constant. Here we use
the power law assumption to solve the integral part in the above
equations. The power law relation between scale factor $a$ and
scalar field $\phi$ has already been used by Johri and Desikan
\cite{18} in the context of Robertson Walker Brans-Dicke models.
Thus the power law relation between $\phi$ and $a$, i.e.
$\phi\propto a^m$, where $m$ is any integer, implies that
\begin{equation}\label{21}
\phi=ba^m,
\end{equation}
where $b$ is the constant of proportionality. The deceleration
parameter $q$ in cosmology is the measure of the cosmic
acceleration of the universe expansion and is defined as
\begin{equation}\label{22}
q=-\frac{\ddot{a}a}{\dot{a}^2}.
\end{equation}
It is mentioned here that $q$ was supposed to be positive
initially but recent observations from the supernova experiments
suggest that it is negative. Thus the behavior of the universe
models depend upon the sign of $q$. The positive deceleration
parameter corresponds to a decelerating model while the negative
value provides inflation. We also use a well-known relation
\cite{19} between the average Hubble parameter $H$ and average
scale factor $a$ given as
\begin{equation}\label{27}
H=la^{-n},
\end{equation}
where $l>0$ and $n\geq0$. This is an important relation because it
gives the constant value of the deceleration parameter. From
Eq.(\ref{60-0008}) and Eq.(\ref{27}), we get
\begin{equation}\label{28}
\dot{a}=la^{-n+1}.
\end{equation}
Using this value, we find that deceleration parameter is constant,
i.e. $q=n-1$. Integrating Eq.(\ref{28}), it follows that
\begin{equation}\label{30}
a=(nlt+k_1)^{\frac{1}{n}},\quad n\neq0
\end{equation}
and
\begin{equation}\label{31}
a=k_2\exp(lt),\quad n=0,
\end{equation}
where $k_1$ and $k_2$ are constants of integration. Thus we obtain
two values of the average scale factor that correspond to two
different models of the universe.

\subsection{Dark Energy Model of the Universe when $n\neq0$.}

Now we discuss the model of universe when $n\neq0$, i.e.,
$a=(nlt+k_1)^{\frac{1}{n}}$. For this model, $\phi$ becomes
\begin{equation}\label{34}
\phi=b(nlt+k_1)^{-\frac{2}{n}}.
\end{equation}
Using this value of $\phi$ in Eq.(\ref{20}), the metric
coefficients $A,~B$ and $C$ turn out to be
\begin{eqnarray}\label{35}
A&=&B={c_2}^\frac{1}{3}(nlt+k_1)^{\frac{1}{n}}\exp[\frac{\lambda(nlt+k_1)^
{\frac{n-1}{n}}}{3b(n-1)}],\quad n\neq1\\\label{36}
C&=&{c_2}^\frac{-2}{3}(nlt+k_1)^{\frac{1}{n}}\exp[\frac{\lambda(nlt+k_1)^
{\frac{n-1}{n}}}{3b(n-1)}],\quad n\neq1.
\end{eqnarray}
The directional Hubble parameters $H_i$ ($i=1,2,3$) take the form
\begin{eqnarray}\label{37}
H_1&=&H_2=\frac{l}{nlt+k_1}+\frac{l\lambda}{3b(nlt+k_1)^{\frac{1}{n}}},\\\label{38}
H_3&=&\frac{l}{nlt+k_1}-\frac{2l\lambda}{3b(nlt+k_1)^{\frac{1}{n}}}.
\end{eqnarray}
The mean generalized Hubble parameter becomes
\begin{equation}\label{39}
H=\frac{l}{nlt+k_1}
\end{equation}
while the volume scale factor turns out to be
\begin{equation}\label{40}
V=(nlt+k_1)^\frac{3}{n}.
\end{equation}
The expansion scalar $\theta$ and shear scalar $\sigma$ take the
form
\begin{eqnarray}\label{41}
\theta&=&\frac{3l}{nlt+k_1},\\
\label{42}
\sigma^2&=&\frac{\lambda^2l^2}{3b^2(nlt+k_1)^{\frac{2}{n}}}
\end{eqnarray}
The mean anisotropy parameter $A_m$ becomes
\begin{equation}\label{43}
A_m=\frac{2\lambda^2}{9b^2}(nlt+k_1)^{2-\frac{2}{n}}.
\end{equation}

Moreover, Eqs.(\ref{14})-(\ref{16}) take the form
\begin{equation}\label{44}
\frac{8\pi\rho}{\phi}=-l^2(3+2\varpi)(nlt+k_1)^{-2}-
[\frac{\lambda^2l^2}{3b^2}+\frac{m^2}{{c_2}^{\frac{2}{3}}}
\exp({\frac{-2\lambda(nlt+k_1)^{\frac{n-1}{n}}}{3b(n-1)}})](nlt+k_1)^{-\frac{2}{n}},
\end{equation}
\begin{equation}\label{45}
-\frac{8\pi\omega\rho}{\phi}=l^2[(3+2n+2\varpi)(nlt+k_1)^{-2}+
\frac{\lambda^2}{3b^2}(nlt+k_1)^{-\frac{2}{n}}],
\end{equation}
\begin{equation}\label{46}
-\frac{8\pi(\omega+\delta)\rho}{\phi}=
l^2(3+2n+2\varpi)(nlt+k_1)^{-2}+[\frac{\lambda^2l^2}{3b^2}-\frac{m^2}{{c_2}^{\frac{2}{3}}}
\exp({\frac{-2\lambda(nlt+k_1)^{\frac{n-1}{n}}}{3b(n-1)}})](nlt+k_1)^{-\frac{2}{n}}.
\end{equation}

\subsection{Dark Energy Model of the Universe when $n=0$.}

The average scale factor for this model of the universe is
$a=k_2\exp(lt)$ and hence $\phi$ takes the form
\begin{equation}\label{49}
\phi=\frac{b}{{k_2}^2}\exp(-2lt).
\end{equation}
Inserting this value of $\phi$ in Eq.(\ref{20}), the metric
coefficients $A,~B$ and $C$ become
\begin{eqnarray}\label{50}
A&=&B={c_2}^{\frac{1}{3}}k_2\exp(lt)\exp[-\frac{\lambda\exp(-lt)}{3blk_2}],\\
\label{51}
C&=&{c_2}^{-\frac{2}{3}}k_2\exp(lt)\exp[\frac{2\lambda\exp(-lt)}{3blk_2}].
\end{eqnarray}
The directional Hubble parameters $H_i$ become
\begin{eqnarray}\label{52}
H_1&=&H_2=l+\frac{\lambda\exp(-lt)}{3bk_2},\\
\label{53} H_3&=&l-\frac{2\lambda\exp(-lt)}{3bk_2},
\end{eqnarray}
while the mean generalized Hubble parameter and the volume scale
factor turn out to be
\begin{equation}\label{54}
H=l,~~~~~~~~~~~~~V={k_2}^3\exp(3lt).
\end{equation}
The expansion scalar $\theta$ and shear scalar $\sigma$ take the
form
\begin{equation}\label{56}
\theta=3l,~~~~~~~~~~~~~
\sigma^2=\frac{\lambda^2\exp(-2lt)}{3b^2{k_2}^2}.
\end{equation}
The mean anisotropy parameter $A_m$ for this model become
\begin{equation}\label{57}
A_m=\frac{2\lambda^2\exp(-2lt)}{9b^2l^2{k_2}^2}.
\end{equation}

For this exponential model of universe, Eqs.(\ref{14})-(\ref{16})
take the form
\begin{equation}\label{440}
\frac{8\pi\rho}{\phi}=-l^2(3+2\varpi)-
\frac{\lambda^2\exp(-2lt)}{3b^2{k_2}^2}-\frac{m^2}{{c_2}^{\frac{2}{3}}{k_2}^2}
\exp(-2lt+{\frac{2\lambda\exp(-lt)}{3blk_2}}),
\end{equation}
\begin{equation}\label{450}
-\frac{8\pi\omega\rho}{\phi}=l^2(3+2\varpi)+
\frac{\lambda^2\exp(-2lt)}{3b^2{k_2}^2},
\end{equation}
\begin{equation}\label{460}
-\frac{8\pi(\omega+\delta)\rho}{\phi}=l^2(3+2\varpi)+
\frac{\lambda^2\exp(-2lt)}{3b^2{k_2}^2}-\frac{m^2}{{c_2}^{\frac{2}{3}}{k_2}^2}
\exp(-2lt+{\frac{2\lambda\exp(-lt)}{3blk_2}}).
\end{equation}

\section{Singularity Analysis}

The Riemann tensor is a useful tool to determine whether a
singularity is essential or coordinate. If the curvature becomes
infinite at a certain point, then the singularity is essential. We
can construct different scalars from the Riemann tensor and thus
it can be verified whether they become infinite somewhere or not.
Infinite many scalars can be constructed from the Riemann tensor,
however, symmetry considerations can be used to show that there
are only a finite number of independent scalars. All others can be
expressed in terms of these. In a four-dimensional Riemann
spacetime, there are only $14$ independent curvature invariants.
Some of these are
\begin{eqnarray*}
R_1=R=g^{ab}R_{ab},\quad R_2=R_{ab}R^{ab},\quad
R_3=R_{abcd}R^{abcd},\quad R_4=R^{ab}_{cd}R^{cd}_{ab}.
\end{eqnarray*}
Here we give the analysis for the first invariant commonly known
as the Ricci scalar for both models.

For model of universe with power law expansion, we can write Ricci
scalar
\begin{equation}\label{329}
R=-2[\frac{l^2(6-3n)}{a^{2n}}+\frac{{\lambda}^2l^2}{3b^2a^2}-
\frac{m^2}{{c_2}^{\frac{2}{3}}a^2\exp({\frac{{2\lambda}a^{n-1}}{3b(n-1)}})}],
\end{equation}
while for exponential model, it is given by
\begin{equation}\label{3293}
R=-2[6l^2+\frac{{\lambda}^2}{3b^2a^2}-
\frac{m^2}{{c_2}^{\frac{2}{3}}a^2\exp({\frac{{2\lambda}a}{3bl}})}].
\end{equation}
Both of these models show that singularity occurs at $a=0$.

\section{Concluding Remarks}

This paper is devoted to explore the solutions of Bianchi type
$III$ cosmological models in Brans Dicke theory of gravitation in
the background of anisotropic dark energy. We use the power law
relation between $\phi$ and $a$ to find the solution. The
assumption of constant deceleration parameter leads to two models
of universe, i.e. power law model and exponential model. Some
important cosmological physical parameters for the solutions such
as expansion scalar $\theta$, shear scalar $\sigma^2$, mean
anisotropy parameter and average Hubble parameter are evaluated.

First we discuss power law model of the universe. This model
corresponds to $n\neq0$ with average scale factor
$a=(nlt+k_1)^{\frac{1}{n}}$. It has a point singularity at
$t\equiv t_s=-\frac{k_1}{nl}$. The physical parameters
$H_1,~H_2,~H_3$ and $H$ are all infinite at this point but the
volume scale factor vanishes here. The metric functions $A,~B$ and
$C$ vanish at this point of singularity. Thus, it is concluded
from these observations that the model starts its expansion with
zero volume at $t=t_s$ and it continues to expand for $0<n<1$.

The exponential model of the universe corresponds to $n=0$ with
average scale factor $a=k_2\exp(lt)$. It is non-singular because
exponential function is never zero and hence there does not exist
any physical singularity for this model. The physical parameters
$H_1,~H_2,~H_3$ are all finite for all finite values of $t$. The
mean generalized Hubble parameter $H$ is constant while metric
functions $A,~B$ and $C$ do not vanish for this model. The volume
scale factor increases exponentially with time which indicates
that the universe starts its expansion with zero volume from
infinite past.

The isotropy condition, i.e., $\frac{\sigma^2}{\theta}\rightarrow
0$ as $t\rightarrow \infty$, is also satisfied in each case. It is
mentioned here that the behavior of these physical parameters is
consistent with the results already obtained in GR \cite{5}. The
variable EoS parameter $\omega$ for both models turn out to be
\begin{equation*}
\omega=\frac{l^2[(3+2n+2\varpi)(nlt+k_1)^{-2}+
\frac{\lambda^2}{3b^2}(nlt+k_1)^{-\frac{2}{n}}]}{l^2(3+2\varpi)(nlt+k_1)^{-2}+
[\frac{\lambda^2l^2}{3b^2}+\frac{m^2}{{c_2}^{\frac{2}{3}}}
\exp({\frac{-2\lambda(nlt+k_1)^{\frac{n-1}{n}}}{3b(n-1)}})](nlt+k_1)^{-\frac{2}{n}}},
\end{equation*}
\begin{equation*}
\omega=\frac{l^2(3+2\varpi)+\frac{\lambda^2\exp(-2lt)}{3b^2{k_2}^2}}
{l^2(3+2\varpi)+
\frac{\lambda^2\exp(-2lt)}{3b^2{k_2}^2}+\frac{m^2}{{c_2}^{\frac{2}{3}}{k_2}^2}
\exp(-2lt+{\frac{2\lambda\exp(-lt)}{3blk_2}})}.
\end{equation*}
Both of these equations suggest that at $t=0$, $\omega$ has a
positive value which indicates that the universe was matter
dominated in its early phase of its existence. At $t\rightarrow
\infty$, the value of $\omega$ turns out to be zero which indicate
that the pressure of the universe vanishes at that epoch.


\vspace{0.05cm}

\begin{thebibliography}{40}

\bibitem{1}Bennett., C.L. et al.: Astrophys. J. Suppl. \textbf{148}(2003)1;
Riess, A.G. et al.: Astrophys. J. \textbf{607}(2004)665; Riess,
A.G. et al. (Supernova Search Team): Astron. J.
\textbf{116}(1998)1009.

\bibitem{2}Spergel, D.N. et al.: Astrophys. J. Suppl. \textbf{148}(2003)175;
ibid. \textbf{170}(2007)377.

\bibitem{3}Tegmartk, M. et al.: Phys. Rev. \textbf{D69}(2004)103501.

\bibitem{04}Riess, A.G. et al. (Supernova Search Team): Astron. J.
\textbf{116}(1998)1009; Perlmutter, S. et al.: Astrophys. J.
\textbf{517}(1999)565; Peebles, P.J.E. and Ratra, B.: Rev. Mod.
Phys. \textbf{75}(2003)559.

\bibitem{004}Jimenez, R.: New astron. Rev. \textbf{47}(2003)761.

\bibitem{4}Knop, R.K. et al.: (Supernova Cosmology Project Collaboration):
Astrophys. J. \textbf{598}(2003)102; Tegmark, M., et al.:
Astrophys. J. \textbf{606}(2004)702.

\bibitem{5}Sharif, M. and Zubair, M.: Int. J. Mod. Phys.
\textbf{D19}(2010)1957.

\bibitem{5001}Sharif, M. and Zubair, M.: Astrophys. Space Sci.
\textbf{330}(2010)399.

\bibitem{5003}Akarsu, O., Kilinc, C.B.: Gen. Rel. Grav.
\textbf{42}(2010)763.

\bibitem{5004}Rahaman, F., Bhui, B., and Bhui, B.C.: Astrophys. Space Sci.
\textbf{301}(2006)47.

\bibitem{5005}Usmani, A.A., Ghosh, P.P.,
Mukhopadhyay, U., Ray, P.C. and Ray, S.: Mon. Not. Roy. Astron.
Soc. Lett. \textbf{386}(2008)L92, arXiv: 0801.4529.

\bibitem{6}Brans, C. and Dicke, R. H.: Phys. Rev. \textbf{124}(1961)925.

\bibitem{614}Lorenz-Petzold, D.: Astrophys. Space Sci.
\textbf{85}(1982)59.

\bibitem{615}Hanquin, J.L. and Demaret, J.: Class. Quantum Grav.
\textbf{1}(1984)291.

\bibitem{616}Tikekar, R. and Patel, L.K.: Gen. Relativ. Grav.
\textbf{24}(1992)397.

\bibitem{617}Yavuz, I. and Yilmaz, I.: Astrophys. Space Sci.
\textbf{245}(1996)131.

\bibitem{14}Sharif, M. and Shamir, M.F.: Class. Quantum Grav.
\textbf{26}(2009)235020.

\bibitem{15}Sharif, M. and Shamir, M.F.: Gen. Relativ. Gravit.
\textbf{42}(2010)2643.

\bibitem{16}Shamir, M.F.: Astrophys. Space Sci.
\textbf{330}(2010)183.

\bibitem{17}Shamir, M.F.: Int. J.
Theor. Phys. \textbf{50}(2011)637.

\bibitem{6017}Chakraborty, S., Chakraborty, N.C. and Debnath, U.:
Int. J. Mod. Phys. \textbf{D12}(2003)325.

\bibitem{6018}Reddy, D.R.K., Adhav, K.S., Katore, S.D. and Wankhade, K.S.:
Int. J. Theor. Phys. \textbf{48}(2009)2884.

\bibitem{618}Christodoulakis, T. and Terzis, P.A.: Class. Quantum Grav.
\textbf{24}(2007)875.

\bibitem{619}Bagora, A.: Astrophys. Space Sci.
\textbf{319}(2009)155.

\bibitem{620}Moussiaux, A., Tombal, P. and Demaret, J.: J. Phys. A: Math. Gen.
\textbf{14}(1981)277.

\bibitem{621}Xing-Xiang, W.: Chin. Phys. Lett.
\textbf{22}(2005)29.

\bibitem{6022}Wang, X.: Astrophys. Space
Sci. \textbf{298}(2005)433.

\bibitem{622}Upadhaya, R.D. and Dave, S.: Braz. J. Phys.
\textbf{38}(2008)4.

\bibitem{6023}Hellaby, C.: Proceedings of Science,
5th International School on Field Theory and Gravitation, April
20-24, 2009, arXiv:0910.0350.

\bibitem{7}Belinskii, V.A. and Khalatnikov, I.M.: Soviet
Physics–JETP, \textbf{36}(1973)591; Reddy, D.R.K. and Rao, V.U.M.:
J. Phys. \textbf{A14}(1981)1973; Banerjee, A. and Santos, N.O.:
Nuovo Cimento \textbf{B67}(1982)31; Ram, S.: Gen. Relativ. Gravit.
\textbf{15}(1983)635.

\bibitem{8}Singh, T. and Rai, L.N.: Astrophys. Space Sci.
\textbf{96}(1983)95.

\bibitem{9}Lorenz-Petzold, D.: Astrophys. Space Sci.
\textbf{85}(1982)59.

\bibitem{10}Kumar, S. and Singh, C.P.: Int. J.
Theor. Phys. \textbf{47}(2008)1722.

\bibitem{11}Adhav, K.S., Ugale,  M.R., Kale, C.B. and Bhende, M.P.:
Int. J. Theor. Phys. \textbf{48}(2009)178.

\bibitem{12}Pradhan, A. and Amirhashchi, H.: Astrophys. Space Sci.
\textbf{332}(2011)441.


\bibitem{13}Adhav, K. S., Nimkar, A.S., Ugale,  M.R. and Dawande, M.V.: Int. J. Theor. Phys.
\textbf{47}(2008)634.

\bibitem{18}Johri, V.B. and Desikan, K.: Gen. Relativ. Grav.
\textbf{26}(1994)1217.

\bibitem{19}Berman, M.S.: IL Nuovo Cim.
\textbf{B74}(1983)182.

\end{thebibliography}
\end{document}